\documentclass[%
 reprint,
 amsmath,amssymb,
 aps,
]{revtex4-1}

\usepackage{amssymb, amsmath}
\usepackage{subfigure}
\usepackage{url}
\usepackage{chemfig}

\begin{document}

\markboth{A. Adamatzky}
{On dynamics of excitation in F-actin}

\title{On dynamics of excitation in F-actin: automaton  model}

\author{Andrew Adamatzky}

\affiliation{University of the West of England,
Bristol BS16 1QT, United Kingdom}
\email{andrew.adamatzky@uwe.ac.uk}

\begin{abstract}
 We represent a filamentous actin molecule as a graph of finite-state machines (F-actin automaton). Each node in the graph takes three states --- resting, excited, refractory. All nodes update their states simultaneously and by the same rule, in discrete time steps. Two rules are considered: threshold rule --- a resting node is excited if it has at least one excited neighbour and narrow excitation interval rule --- a resting node is excited if it has exactly one excited neighbour. We analyse distributions of transient periods and lengths of limit cycles in evolution of F-actin automaton, propose mechanisms for formation of limit cycles and evaluate density of information storage in F-actin automata.   
\end{abstract}

 \keywords{actin; excitation; automata; dynamics; computation}

 \maketitle

\graphicspath{{figs/}}

\section{Introduction}

\begin{figure}[!bp]
    \centering
    \includegraphics[width=1\linewidth]{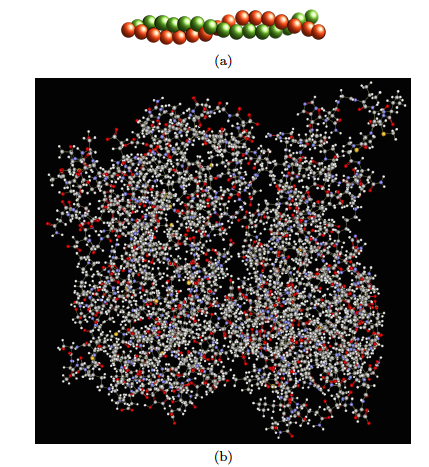}
    \caption{Actin structure.
    (a) Schematic drawing an actin helix. Spheres are coloured for visual guidance only. Each sphere is a F-actin unit, which molecular structures is shown in (b).
    (b) F-actin molecule, CPK colouring.
    }
    \label{fig:actinballstick}
\end{figure}

Actin is a  protein presented in all eukaryotic cells in the forms of globular actin (G-actin) and filamentous actin (F-actin)~\cite{straub1943actin,korn1982actin,szent2004early}.  
G-actin, polymerises in double helix of filamentous actin  (Fig.~\ref{fig:actinballstick}a), during polymerisation G-actin units slightly change their shapes and thus become F-actin units~\cite{oda2009nature}. The actin filaments form a skeleton of single cells, where they play key roles in motility and shape changing --- together myosin --- and signal transduction -- together with tubulin microtubules~\cite{cooper2000cell}.  Actin filaments networks are key components of neural synapses~\cite{cingolani2008actin}. The actin networks is a substrate of cell-level learning~\cite{hameroff1988coherence,rasmussen1990computational,ludin1993neuronal,conrad1996cross,tuszynski1998dielectric,priel2006dendritic,debanne2004information,priel2010neural,jaeken2007new} and information processing~\cite{fifkova1982cytoplasmic,kim1999role,dillon2005actin,cingolani2008actin}. Actin filaments process information in synapses and cells, they compute in a hardwired sense, as specialised processors. If we did manage to uncover exact mechanisms of information transmission and processing in the actin filaments and establish an interface with actin filaments we would be able to make a large-scale massive-parallel nano-computing devices. In ~\cite{adamatzky2015actin} we proposed a model of  actin filaments as two chains of one-dimensional binary-state semi-totalistic automaton arrays. We discovered automaton state transitions rules that support travelling localisations, compact clusters of non-resting states. These travelling localisations are analogous to  ionic waves proposed in actin filaments~\cite{tuszynski2004ionic}. We speculated that a computation in actin filaments could be implemented when localisations (defects, conformation changes, ionic clouds, solitons), which represent data, collide with each other and change their velocity vectors or states.
Parameters of the localisations before a collision are interpreted as values of input variables. Parameters of the localisation after the collision are values of output variables. We implemented a range of computing schemes in several families of actin filament models, from quantum automata to lattice with Morse potential~\cite{siccardi2015actin,siccardi2016boolean,siccardi2016quantum,siccardi2016logical,siccardi2017models}. 
These models considered a unit (F-actin) of an actin filament as a single, discrete, entity which can take just two or three states, and carriers of information occupied one-two actin units. These were models of rather coarse-grained computation~\cite{siccardi2015actin,siccardi2016boolean,siccardi2016quantum,siccardi2016logical,siccardi2017models}. To take the paradigm of computation via interaction travelling localisations at the sub-molecular level we must understand how information, presented by a perturbation of some part of an F-actin unit from its resting state, propagates in the F-actin unit. The paper is structured as follows. We define a model of F-actin automata in Sect.~\ref{model}. In Sect.~\ref{dynamicsA0} we study excitation dynamics of automata with a threshold excitation rule, and in Sect.~\ref{dynamicsA1} with a rule of narrow excitation interval. Implications of our findings for designs of actin-based information storage  devices are discussed in Sect.~\ref{discussion}.

\section{Model}
\label{model}

We use a   structure of F-actin molecule  produced  using X-ray fibre diffraction intensities obtained  from well oriented sols of rabbit skeletal muscles~\cite{oda2009nature}. The structure 
was calculated with resolution 3.3\AA~in radial direction and 5.6\AA~along the axis (Fig.~\ref{fig:actinballstick}b)~\cite{oda2009nature}.
The molecular structure was converted to a non-directed graph $\mathcal A$, where every node represents an atom and  an edge corresponds to a bond between the atoms. 
The  graph $\mathcal A$ has 2961 nodes, 3025 edges. Minimum degree is 1, maximum is 4, average is 2.044 (with standard deviation 0.8) and median degree 2. There are 883 nodes with degree 1, 1009 nodes with degree 2, 1066 nodes with degree 3 and two nodes with degree 4. The graph $\mathcal A$ has a diameter (longest shortest path) 1130 nodes, and a mean distance (mean shortest path between any two nodes) 376, and a median distance 338. 

We study dynamics of excitation in the actin graph $\mathcal A$ using the following models. Each node $s$ of $\mathcal A$ takes three states: resting ($\circ$), excited ($\oplus$) and refractory ($\ominus$). Each node $s$ has a neighbourhood $u(s)$ which is a set of nodes connected to the node $s$ by edges in $\mathcal A$. The nodes update their states simultaneously in discrete time by the same rule. Each step of simulated discrete time corresponds to one attosecond of real time. 

A resting node $s^t=\circ$ excites depending on a number  $\sigma_s^t$ of its excited neighbours in neighbourhood $u(s)$:  $\sigma_s^t = \sum_{w \in u(s)} \{w^t=\oplus \}$ We consider two excitation rules.
 In rule ${\mathcal A}_0$  a resting node excites if it has at least one excited neighbour:  $\sigma_s^t>0$. 
 In rule ${\mathcal A}_1$  a resting node excites if it has exactly one excited neighbour: $\sigma_s^t=1$ (we do not consider rules where $\sigma_s^t>1$ because excitation there extincts quickly). Transitions from excited state to refractory state and from refractory state to resting state are unconditional, i.e. these transitions take place independently on neighbourhood state.
 The rules can be written as follows\\
\begin{tabular}{ccc}
 ${\mathcal A}_0$  &  & ${\mathcal A}_1$  \\
 $
s^{t+1}=
\begin{cases}
\oplus, \text{ if } \sigma_s^t>0 \\
\ominus, \text{ if } s^t=\oplus\\
\circ, \text{ otherwise }
\end{cases}
$ 
& &
$
s^{t+1}=
\begin{cases}
\oplus, \text{ if } \sigma_s^t=1 \\
\ominus, \text{ if } s^t=\oplus\\
\circ, \text{ otherwise }
\end{cases}
$
\end{tabular}\\

At the beginning of each computational experiment the F-actin automaton $\mathcal A$ is in a global resting state, every node is assigned state $\circ$. 
An excitation dynamic   is initiated by assigning  non-resting states $\oplus$ or $\ominus$ or both to a portion of randomly selected nodes. 
Three stimulation scenarios are considered:
\begin{itemize}
\item Single node stimulation: a single node is selected at random and this node is assigned excited state $\oplus$
\item $(+)$-stimulation:  a specified ratio of nodes is selected at random and the selected nodes are assigned excited state $\oplus$
\item $(+-)$-stimulation:  a specified ratio of nodes is selected at random and the selected nodes are assigned either excited state $\oplus$ or refractory state $\ominus$ at random.
\end{itemize} 
The automaton $\mathcal A$ is deterministic, therefore from any initial configuration the automaton evolves into in a limit cycle in its state space (where its configuration is repeated after a finite number of steps) or an a absorbing state (this is limit cycle length one). For the rules selected there is only one absorbing state --- all nodes are in the resting state. A limit cycle is comprised of configurations where compact patterns of excitation travel along closed paths.  A transient period is an interval of automaton evolution from initial configuration to entering a limit cycle or an absorbing state. 

For modelling we used C and Processing, for visualisation and analyses we used R, iGraph, Chimera.

\section{Dynamics of  ${\mathcal A}_0$}
\label{dynamicsA0}

\subsection{Single node stimulation}

\graphicspath{{figs/}}

\begin{figure}[!tbp] 
    \centering
    \includegraphics[width=\linewidth]{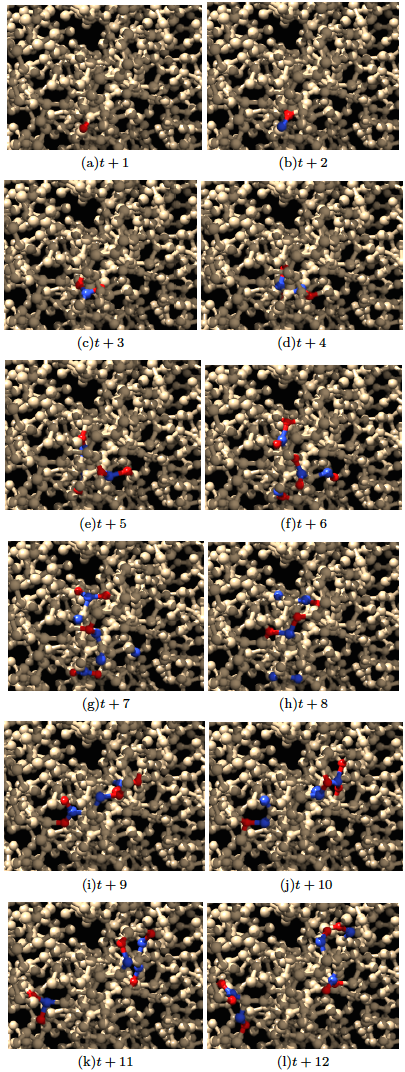}
    \caption{Exemplar excitation dynamics of ${\mathcal A}_0$ in a scenario of a single node stimulation. In initial configuration all nodes are resting but one node is excited.
    (a--f)~Snapshots of the simulation. Excited nodes are red, refractory nodes are blue, resting nodes are light-grey.
    }
    \label{fig:singlenoderuleR0}
\end{figure}


\begin{figure}[!tbp] 
    \centering
    \subfigure[]{\includegraphics[width=0.99\linewidth]{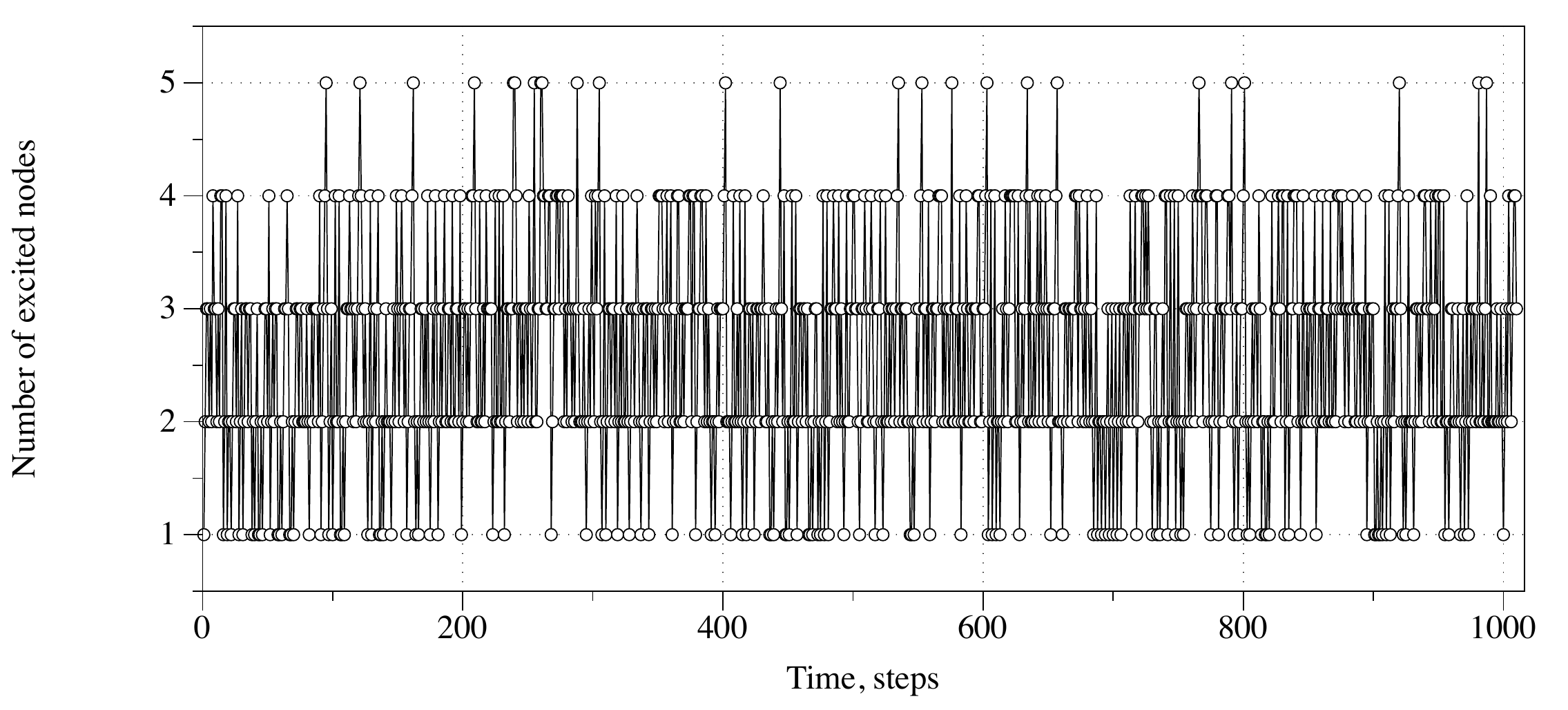}}
    \subfigure[]{\includegraphics[width=0.99\linewidth]{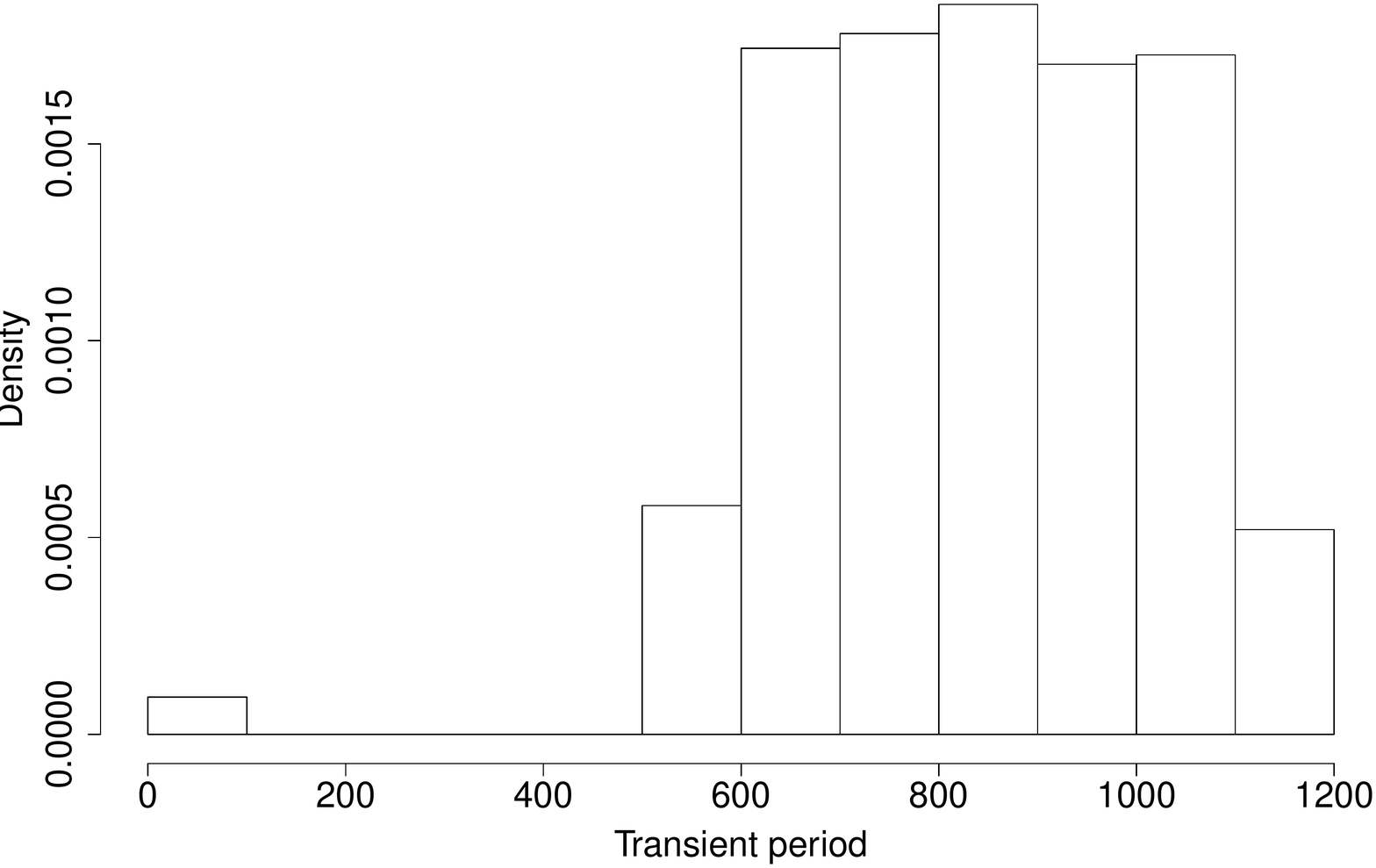}}
    \caption{Integral dynamics of rule ${\mathcal A}_0$ automata in scenarios of  a single node stimulation.
    (a)~Number of excited nodes at each step of simulation in a single experiment.
    (b)~Distribution of transient periods.}
    \label{fig:distributionsinglenodeexcitaiton}
\end{figure} 

\begin{figure}[!tbp]   
\centering
 \includegraphics[width=0.49\textwidth]{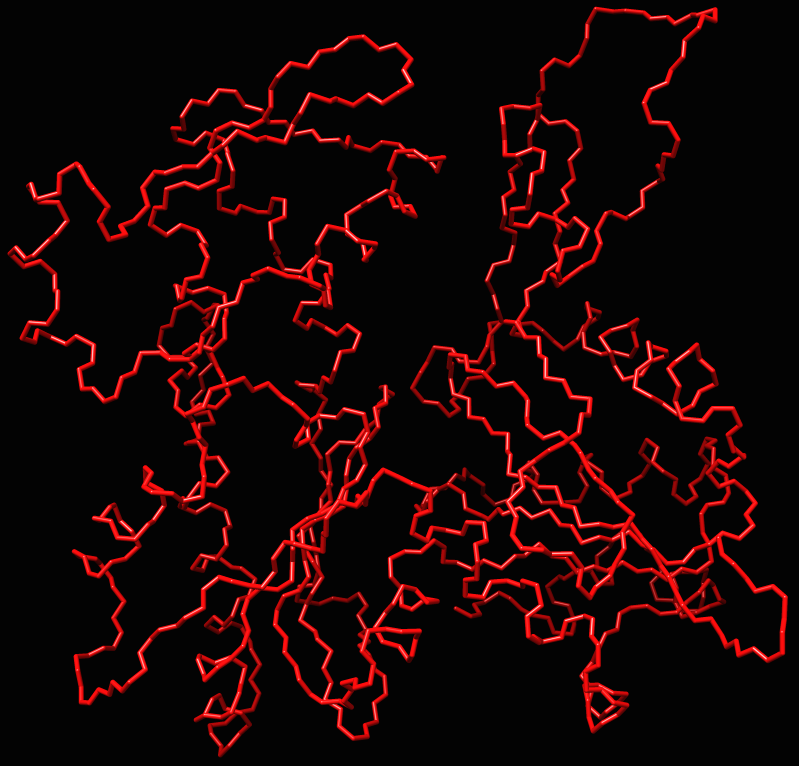}
    \caption{Longest path of excitation propagation in ${\mathcal A}_0$ displayed by bonds.}
    \label{fig:longestexcitation}
\end{figure}

The excitation propagates as a localised pattern  (Fig.~\ref{fig:singlenoderuleR0}a-f). A number of nodes excited at every single step 
of time varies between one and five (Fig.~\ref{fig:distributionsinglenodeexcitaiton}a). Sometimes an excitation pattern splits into two localisations which travel along their independent pathways.
The automaton ${\mathcal A}_0$ always evolves into the absorbing state where all nodes are resting. This is because travelling localisation either cancel each other when collide or reach cul-de-sacs of their
pathways.  A distribution of  transition periods is shown in Fig.~\ref{fig:distributionsinglenodeexcitaiton}b. The mean transition period is 840 time steps, median 847, minimum 2 and maximum 1131. Only 29 nodes, when stimulated lead to excitation development with a transition period between 2 and 15 steps. Stimulation of all others  2932 trigger excitation dynamics for  at least 568 steps. The longest transition period is observed when the localised excitation runs along a longest shortest path where initially stimulated node is a source. The path of the longest excitation is visualised in  Fig.~\ref{fig:longestexcitation}a; the path matches the backbone  of the actin unit.

\subsection{$(+)$-stimulation}

\graphicspath{{figs/}}

\begin{figure}[!tbp] 
    \centering
\includegraphics[width=\linewidth]{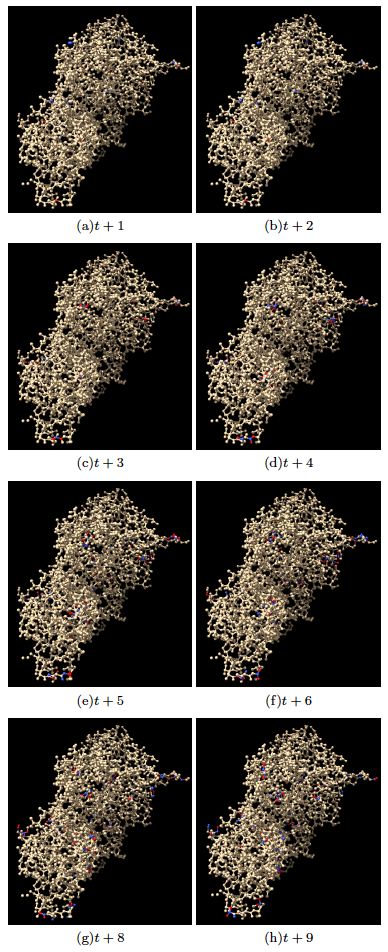}
    \caption{Exemplar dynamics of ${\mathcal A}_1$. In initially resting configuration 1\% of nodes is excited. Edges of $\mathcal A$ are shown by grey colour, excited nodes are red, refractory nodes are blue, resting nodes are not shown.}
    \label{fig:singlenodesnapshots}
\end{figure}

\graphicspath{{figs/}}


\begin{figure}[!tbp] 
    \centering
      \subfigure[]{\includegraphics[width=0.99\linewidth]{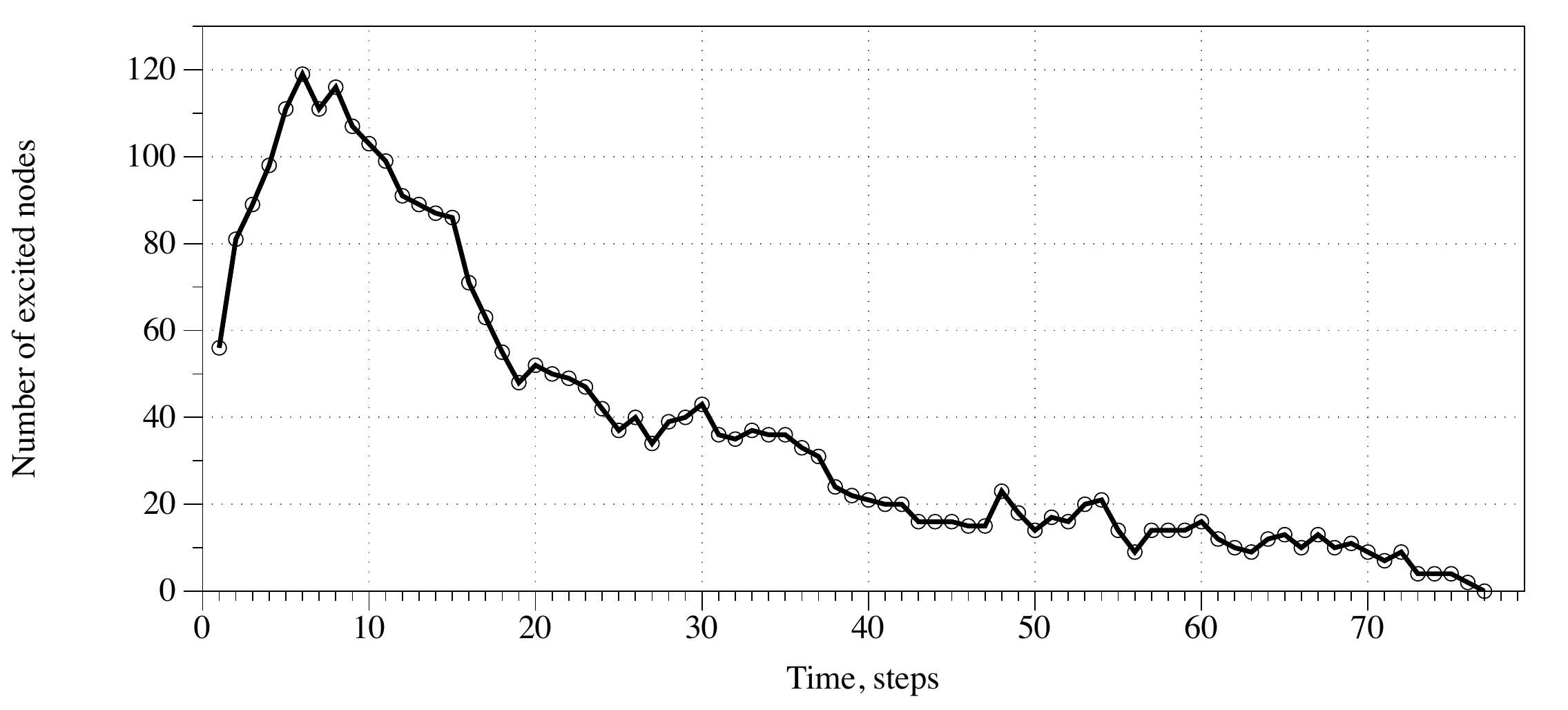}}
      \subfigure[]{\includegraphics[width=0.99\linewidth]{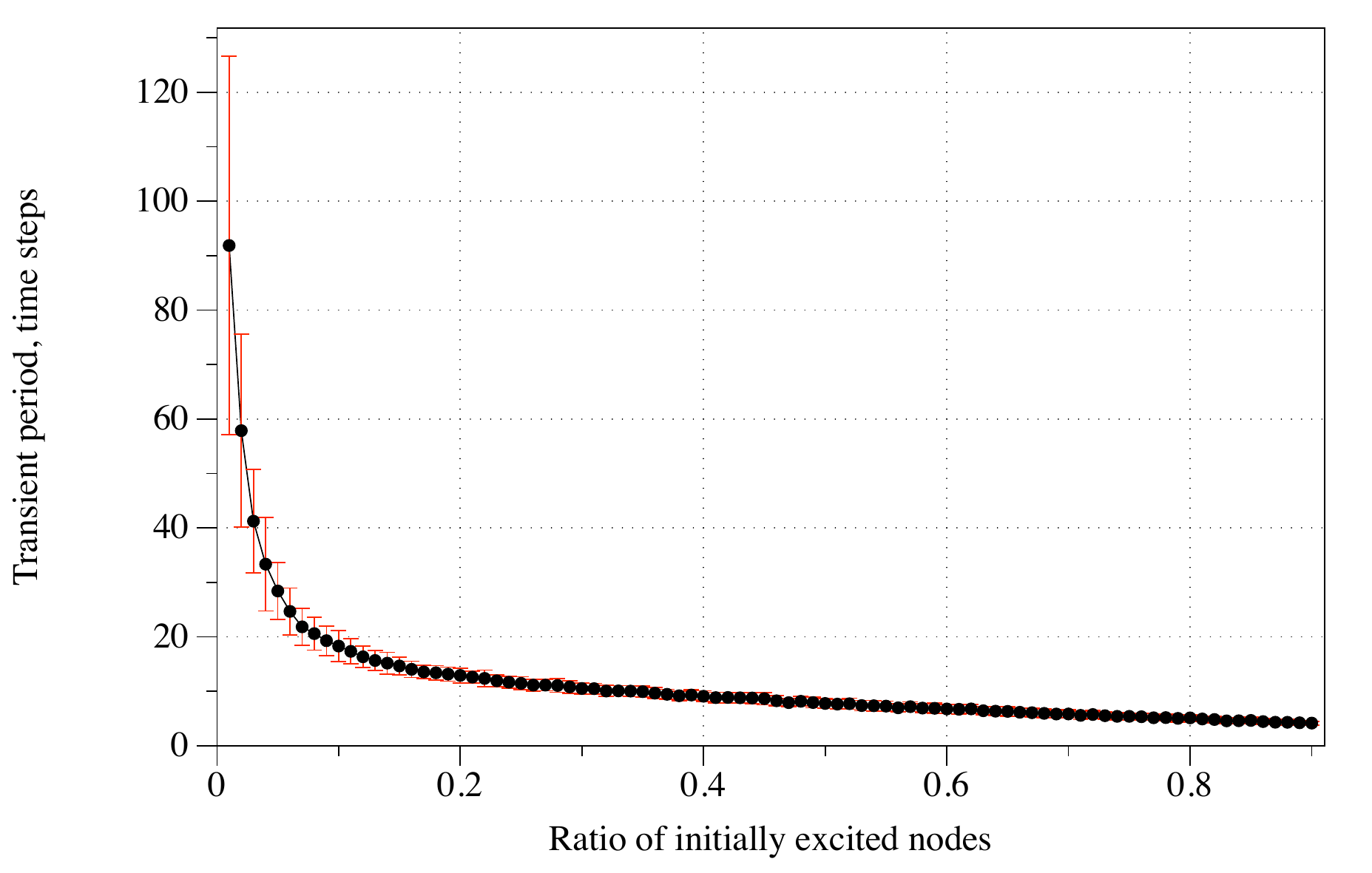}}
    \caption{Integral dynamics of ${\mathcal A}_0$ in scenarios when a portion of nodes is excited initially.  
    (a)~A number of excited nodes versus time: initially, 1\% of nodes are assigned excited states. 
    (b)~Ratio of initially excited nodes versus average transient period. Standard deviation is shown as error bars.}
    \label{fig:excitationdynamicsportionexcited}
\end{figure}

When we stimulate more than one node the automaton ${\mathcal A}_0$ exhibits several `epicentres' of excitation, the patterns of excitation propagate away from their origins (Fig.~\ref{fig:singlenodesnapshots}), and populate the graph. This stage is manifested in increasing a number of excited states at each step of the evolution (Fig.~\ref{fig:excitationdynamicsportionexcited}a). Eventually, depending on distances between sources of excitation, the graph becomes filled with waves and localisations, e.g. in illustration Fig.~\ref{fig:excitationdynamicsportionexcited}a a peak is reached in 7-8 steps.
Then patterns of excitation start colliding with each other. They  annihilate in the results of the collisions. A number of excited nodes decreases over time (Fig.~\ref{fig:excitationdynamicsportionexcited}a). The graph returns to the totally resting state. The larger is the portion of initially excited nodes the quicker evolution halts in the resting state (Fig.~\ref{fig:excitationdynamicsportionexcited}b). The `quicker' can be quantified by a polynomial function $p = 4.7 \cdot \rho^{-0.6}$, where $p$ is a length of transient  period and $\rho$ is a ratio of initially excited nodes.

\subsection{$(+-)$-stimulation}

\begin{figure}[!tbp] 
    \centering
    \includegraphics[width=1\linewidth]{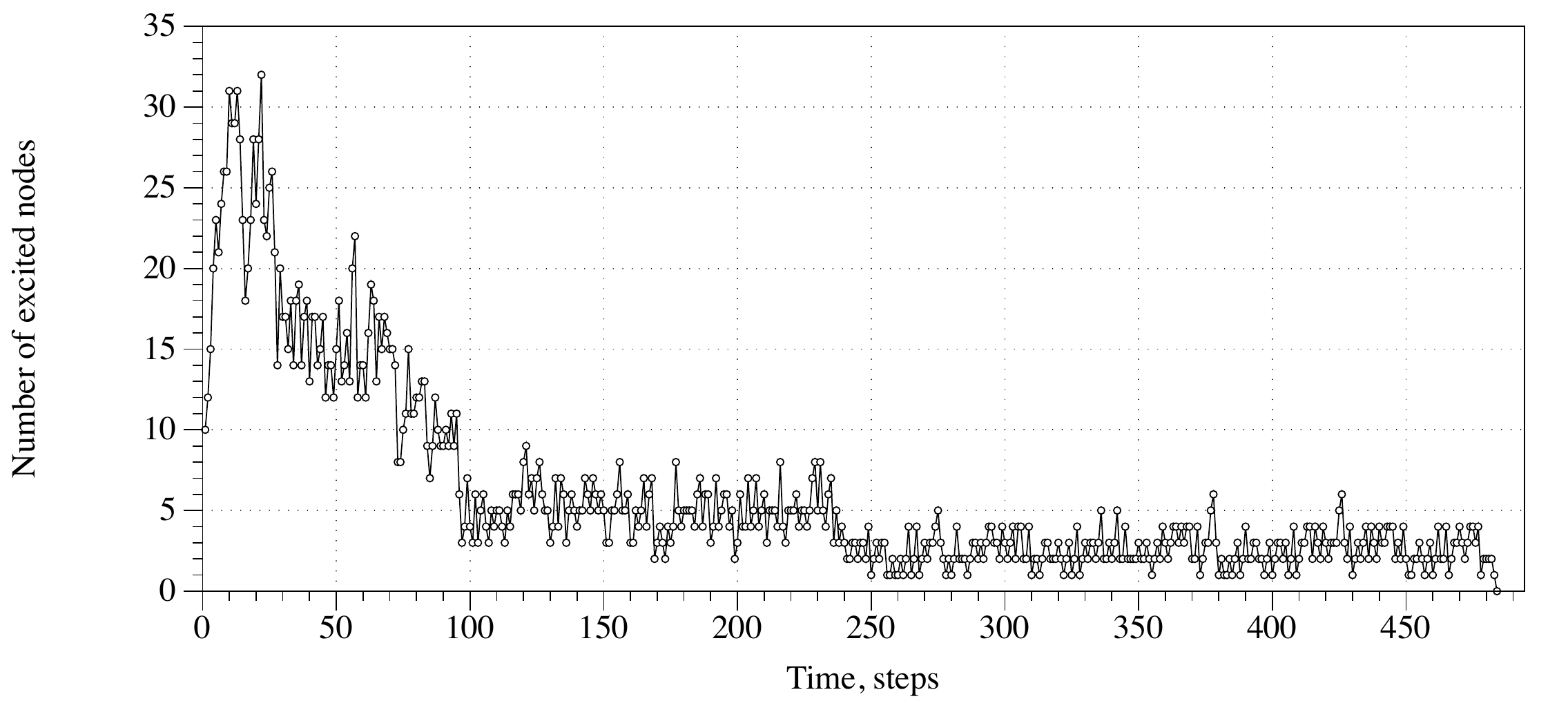}
    \caption{Integral dynamics of ${\mathcal A}_0$ in scenarios when  
    1\% of nodes are assigned excited or refractory states initially.}
    \label{fig:ExcitationDynamocsOnePercentExcRef}
\end{figure} 

In a 'classical' two-dimensional discrete excitable medium stimulation of the medium with an excited node neighbouring with a refractory node leads to formation of a spiral wave. Due to the spiral waves excitation can persist in a modelled medium indefinitely. F-actin automata follow this principle. When we stimulate nodes of ${\mathcal A}_0$ such that some of the nodes get excited states and some refractory states we evoke the excitation patterns. Average level of excitation over trials is proportional to a number of nodes stimulated (see row $e$ in Tab.~\ref{tab:statistics}a).
The automaton enters a  limit cycle (Fig.~\ref{fig:ExcitationDynamocsOnePercentExcRef}). The limit cycle's length varies between 5 to 14 time steps (see row $c$ in Tab.~\ref{tab:statistics}a). Apparently, the automaton falls into longest limit cycles when nearly half of nodes are stimulated, however, due to high deviation of the results (see row $\sigma(c)$ in Tab.~\ref{tab:statistics}a), we would not state this as a fact. Lengths of transient periods, from stimulation to entering the limit cycle, is over a half of the number of nodes in $\mathcal A$.

\section{Dynamics of ${\mathcal A}_1$}
\label{dynamicsA1}

\subsection{Single node stimulation} When a single node is excited initially, the automaton ${\mathcal A}_1$ always evolves to a globally resting state. In sampling of seventy trials we found that average length of 
the transient period is 862 time steps (standard deviation 230) and median transition period is 869. The average transient period to the resting state  
is 22 steps longer then the one in the automaton ${\mathcal A}_0$.


\begin{table*}[!tbp]
\caption{Characterisation of excitation dynamics in  actin automata $\mathcal A$. 
    (abc)~Dependence of the dynamic on ratios of stimulated nodes.
    (a)~${\mathcal A}_0$, initially excited ratio $\rho$ of nodes.
    (b)~${\mathcal A}_1$, initially excited ratio $\rho$ of nodes. 
    (c)~${\mathcal A}_1$, initial ratio $\rho$ of nodes assigned excited and refractory states at random.
    Data is collected in ten experiments  for each value of $\rho$. For ten ratios $\rho$ of initially stimulated nodes, $\rho=0.1, 0.2, \ldots, 0.9$ we calculated lengths of transient period $p$, 
    lengths of cycles $c$, numbers $e$ of excited nodes in a cycle, and their standard deviations $\sigma(p)$, $\sigma(c)$, $\sigma(e)$. Values are rounded to integer.
    (d)~Characteristics averaged all over stimulation ratios for each rule and stimulation scenario.}
    \subtable[${\mathcal A}_0$, $(+-)$-start]{
    \begin{tabular}{c|cccccc}
      $\rho$	&	$p$	&	$c$	&	$e$	&	$\sigma(p)$	&	$\sigma(c)$	&	$\sigma(e)$	\\ \hline
10	&	1613	&	6	&	535	&	1820	&	1	&	55	\\ 
20	&	1432	&	5	&	562	&	988	&	0	&	52	\\ 
30	&	1984	&	7	&	626	&	1275	&	8	&	108	\\ 
40	&	2536	&	14	&	598	&	3064	&	15	&	15	\\ 
50	&	1583	&	13	&	786	&	610	&	8	&	206	\\ 
60	&	2614	&	14	&	719	&	3322	&	14	&	191	\\ 
70	&	2052	&	9	&	805	&	2236	&	4	&	207	\\ 
80	&	1311	&	5	&	705	&	521	&	1	&	180	\\ 
90	&	2850	&	16	&	651	&	2835	&	15	&	157	\\ 
    \end{tabular}}
    \subtable[${\mathcal A}_1$, $(+)$-start]{
    \begin{tabular}{c|cccccc}
      $\rho$	&	$p$	&	$c$	&	$e$	&	$\sigma(p)$	&	$\sigma(c)$	&	$\sigma(e)$	\\ \hline
0.1	&	1154	&	13	&	570	&	1251	&	12	&	39	\\
0.2	&	1388	&	13	&	553	&	961	&	12	&	50	\\
0.3	&	893	&	11	&	575	&	362	&	10	&	8	\\
0.4	&	920	&	24	&	590	&	487	&	11	&	51	\\
0.5	&	996	&	16	&	575	&	832	&	13	&	11	\\
0.6	&	746	&	16	&	594	&	238	&	12	&	24	\\
0.7	&	891	&	16	&	625	&	455	&	11	&	77	\\
0.8	&	1354	&	16	&	639	&	408	&	12	&	109	\\
0.9	&	1729	&	15	&	577	&	1368	&	13	&	13	\\
    \end{tabular}}
     \subtable[${\mathcal A}_1$, $(+-)$-start]{   
    \begin{tabular}{c|cccccc}
$\rho$	&	$p$	&	$c$	&	$e$	&	$\sigma(p)$	&	$\sigma(c)$	&	$\sigma(e)$	\\ \hline
0.1	&	893	    &	118	&	496	&	934	&	355	&	172	\\
0.2	&	2328	&	5	&	583	&	1525	&	0	&	6	\\
0.3	&	1953	&	13	&	599	&	1791	&	12	&	16	\\
0.4	&	1957	&	8	&	591	&	2400	&	8	&	5	\\
0.5	&	1785	&	14	&	636	&	2567	&	12	&	130	\\
0.6	&	976	    &	6	&	706	&	345	&	3	&	179	\\
0.7	&	1342	&	11	&	709	&	464	&	13	&	175	\\
0.8	&	1170	&	13	&	625	&	620	&	14	&	109	\\
0.9	&	2599	&	5	&	593	&	2452	&	1	&	14	\\
    \end{tabular}}
        \subtable[]{
     \begin{tabular}{l|cc|cc}
        &   \multicolumn{2}{c}{${\mathcal A}_0$} &  \multicolumn{2}{c}{${\mathcal A}_1$}  \\ 
        &   $(+)$-start &  $(+-)$-start &  $(+)$-start & $(+-)$-start \\ \hline
 $p$     &       840             &       1997                & 1118          & 1667 \\
 $c$     &           1           &       10                  &   15          &  21 \\
 $e$     &           0           &       665                 &   588         & 615 
    \end{tabular}}
    \label{tab:statistics}
\end{table*}

\subsection{(+)-stimulation} In contrast to automaton  ${\mathcal A}_0$, automaton ${\mathcal A}_1$   does not show a pronounced sensitivity to a ratio $\rho$ of initially excited nodes. Transition periods for all values of $\rho$ are grouped around 1112 (Tab.~\ref{tab:statistics}b). The automata always evolve to limit cycles. Cycle lengths are around 15 time steps with excitation level (number of excited nodes) of just below 600 nodes. The system shows a high degree of variability in lengths of transition periods and cycles, as manifested in large values of standard deviations $\sigma(p)$ and $\sigma(c)$. 
Level of excitation typically remains preserved.

\subsection{(+-)-stimulation}
 
${\mathcal A}_1$  behaves similarly to the scenario of $(+)$-start: there are many travelling localisation, which collide and, mostly, annihilate each other. Few  localisations survive by finding a cyclic path to travel: if no other localisation enters their path, the remaining localisations can cycle indefinitely. The surviving localisations are responsible for  ${\mathcal A}_1$ falling into the limit cycle. Automaton starting with a mix of randomly excited and refractory states usually travels one-and-half times longer to its limit cycle than the same automaton starting only with randomly excited states (compare 
Tab.~\ref{tab:statistics}b and Tab.~\ref{tab:statistics}c).

\section{Stability of the dynamics}

\begin{figure}[!tbp] 
    \centering
   \includegraphics[width=\linewidth]{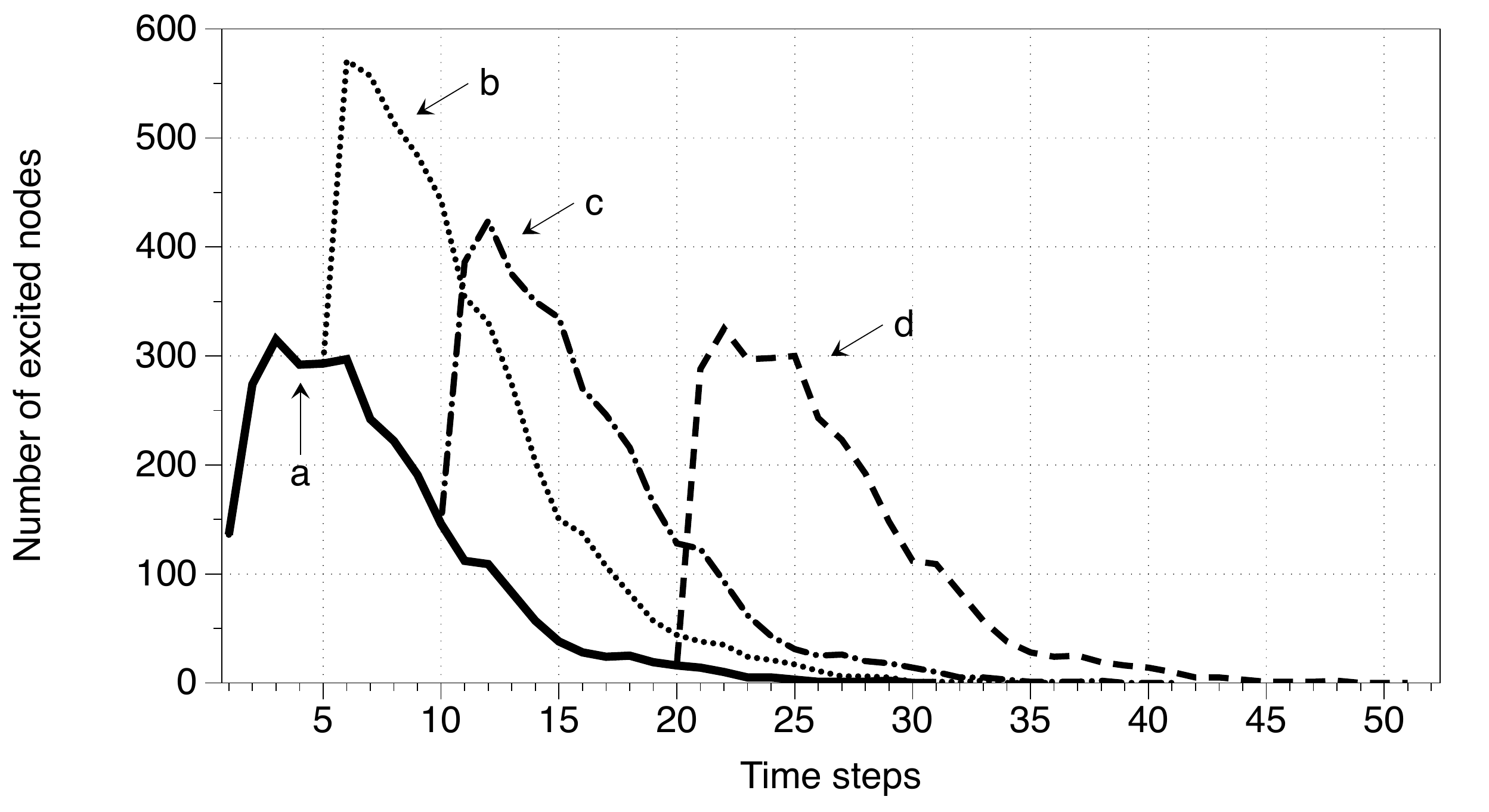}
    \caption{Dynamics of ${\mathcal A}_0$ automaton under repeated stimulation.
    In each trial 5\% of nodes were initially excited.
    (a)~No more stimulation applied (apart of initial stimulation).
    (bcd)~Automaton was stimulated by exciting 5\% of nodes at 5th step (b), 10th step (c) and 20th step (d) steps of evolution.}
    \label{fig:RepeatedExcitation}
\end{figure} 

\begin{figure}[!tbp] 
    \centering
   \includegraphics[width=\linewidth]{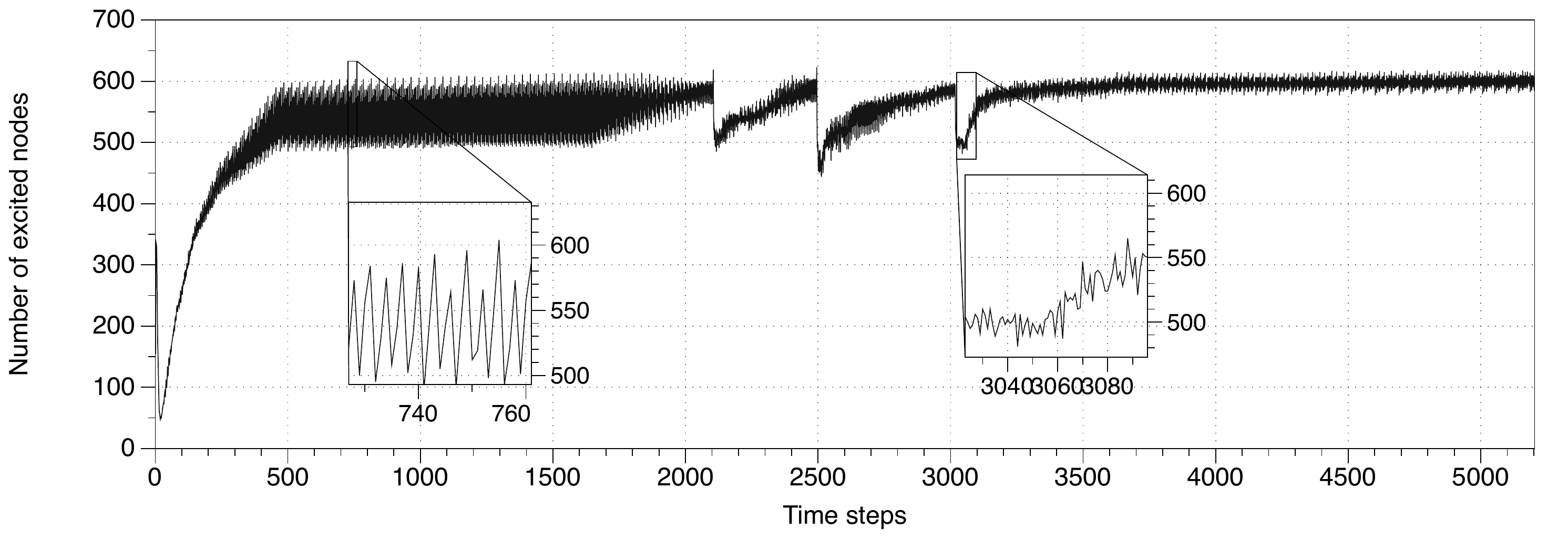}
    \caption{Dynamics of ${\mathcal A}_0$ automaton under repeated stimulation.
    Initially 10\% of nodes were assigned excited or refractory state at random. When the automaton reached limit cycle the stimulation was repeated. }
    \label{fig:RepeatedExcRefWhenCycleReached}
\end{figure} 

How does repeated stimulation affect excitation dynamics of  ${\mathcal A}_0$ and ${\mathcal A}_1$? $(+)$-stimulation of ${\mathcal A}_0$ at any stage of its evolution raises level of excitation by amount equivalent to that of  stimulated resting automaton (Fig.~\ref{fig:RepeatedExcitation}). Thus repeated stimulation prolongs return of the automaton to its resting state. In scenario of $(+-)$-stimulation ${\mathcal A}_0$ evolves to a limit cycle. Repeated $(+-)$-stimulation of the automaton while it is in the limit cycle causes the automaton to change its trajectory in a state space. This change is characterised by initially reduced level of excitation. Typically, excitation level drops by 100-150 nodes at the moment of stimulation. The level of excitation returns to its `pre-stimulation' value in 400-500 time steps.

\section{Implementation of memory}
\label{memory}

\begin{figure*}[!tbp]
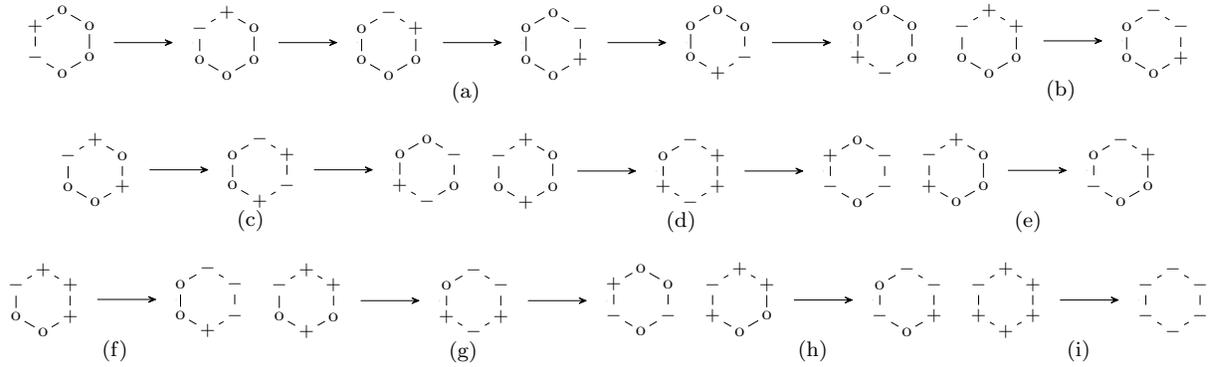

\setatomsep{1.6em}
\subfigure[]{
\scalebox{0.8}{
\schemestart
\chemfig{{-}*6(-o-o-o-o-{+}-)} \arrow 
\chemfig{o*6(-o-o-o-{+}-{-}-)} \arrow
\chemfig{o*6(-o-o-{+}-{-}-o-)} \arrow
\chemfig{o*6(-o-{+}-{-}-o-o-)} \arrow
\chemfig{o*6(-{+}-{-}-o-o-o-)} \arrow
\chemfig{{+}*6(-{-}-o-o-o-o-)} 
\schemestop
}
}
\subfigure[]{
\scalebox{0.8}{
\schemestart
\chemfig{o*6(-o-o-{+}-{+}-{-}-)}\arrow
\chemfig{o*6(-o-{+}-{-}-{-}-o-)} 
\schemestop
}}
\subfigure[]{
\scalebox{0.8}{
\schemestart
\chemfig{o*6(-o-{+}-o-{+}-{-}-)} \arrow
\chemfig{o*6(-{+}-{-}-{+}-{-}-o-)} \arrow
\chemfig{{+}*6(-{-}-o-{-}-o-o-)} 
\schemestop
}}
\subfigure[]{
\scalebox{0.8}{
\schemestart
\chemfig{o*6(-{+}-o-o-{+}-{-}-)} \arrow
\chemfig{{+}*6(-{-}-{+}-{+}-{-}-o-)} \arrow
\chemfig{{-}*6(-o-{-}-{-}-o-+-)}
\schemestop
}}
\subfigure[]{
\scalebox{0.8}{
\schemestart
\chemfig{{+}*6(-o-o-o-{+}-{-}-)} \arrow
\chemfig{{-}*6(-o-o-{+}-{-}-o-)}
\schemestop
}}
\subfigure[]{
\scalebox{0.8}{
\schemestart
\chemfig{o*6(-o-{+}-{+}-{+}-{-}-)} \arrow
\chemfig{o*6(-{+}-{-}-{-}-{-}-o-)} 
\schemestop
}}
\subfigure[]{
\scalebox{0.8}{
\schemestart
\chemfig{o*6(-{+}-o-{+}-{+}-{-}-)}\arrow
\chemfig{{+}*6(-{-}-{+}-{-}-{-}-o-)}\arrow
\chemfig{{-}*6(-o-{-}-o-o-{+}-)}
\schemestop
}}
\subfigure[]{
\scalebox{0.8}{
\schemestart
\chemfig{{+}*6(-o-o-{+}-{+}-{-}-)} \arrow
\chemfig{{-}*6(-o-{+}-{-}-{-}-o-)}
\schemestop
}}
\subfigure[]{
\scalebox{0.8}{
\schemestart
\chemfig{{+}*6(-{+}-{+}-{+}-{+}-{-}-)}\arrow
\chemfig{{-}*6(-{-}-{-}-{-}-{-}-{-}-)}
\schemestop
}}
\caption{Excitation dynamics of aromatic ring automaton governed by ${\mathcal A}_0$.
(a)~Propagation of excitation wave on undisturbed ring automaton.
(b--i)~Stimulation of ring automaton.}
\label{ringscheme}
\end{figure*}

F-actin automata entering limit cycles could play a role of information storage in actin filaments. The minimal length of a limit cycle detected is 5 time steps. Thus aromatic rings could be a substrate responsible for some patterns of cycling excitation dynamics. Let an aromatic ring automaton be stimulated such that a node is assigned an excited state and one of its neighbours is assigned refractory state. The wave of excitation (comprised of one excited and one refractory states) propagates into the direction of its excited head (Fig.~\ref{ringscheme}a).  The excitation running along the aromatic ring can not be extinguished by stimulation of one resting node (Fig.~\ref{ringscheme}bcde) or two resting nodes (Fig.~\ref{ringscheme}fgh). This is because an excited node surrounded by two resting neighbours excites both resting neighbours. Thus excitation waves propagate along the ring in both direction. Therefore even if original excitation wave is cancelled by external stimulation then similar running wave will emerge.  To extinguish the excitation in an aromatic ring we must externally excite all four resting nodes or force them into a refractory state.

\begin{figure}[!tbp] 
    \centering
    \includegraphics[width=\linewidth]{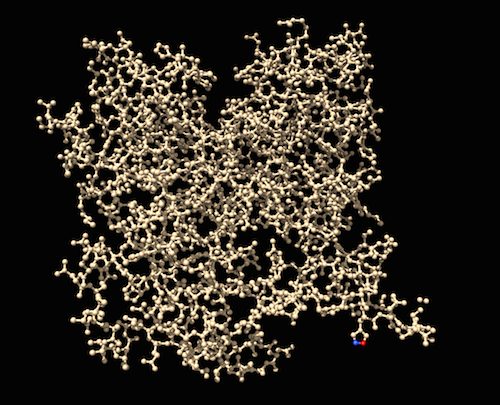}
    \caption{Configuration of resting ${\mathcal A}_0$ at the moment of external stimulation of histidine's aromatic ring. Resting nodes are light-grey, excited nodes are red, refractory nodes are blue. }
    \label{aromatic1}
\end{figure}

\begin{figure}[!tbp] 
    \centering
    \includegraphics[width=\linewidth]{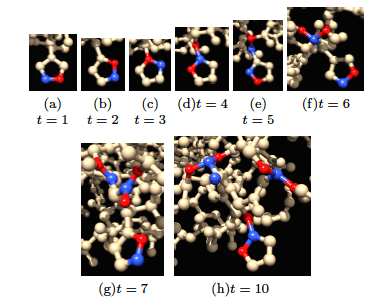}
        \caption{Excitation dynamics of ${\mathcal A}_0$ triggered by excitation of histidine's aromatic ring. First ten steps of the automaton evolution are shown. Only part of the graph adjacent to excitation is displayed. Resting nodes are light-grey, excited nodes are red, refractory nodes are blue.}
    \label{aromatic2}
\end{figure}

\begin{figure}[!tbp]
    \centering
 \includegraphics[width=\linewidth]{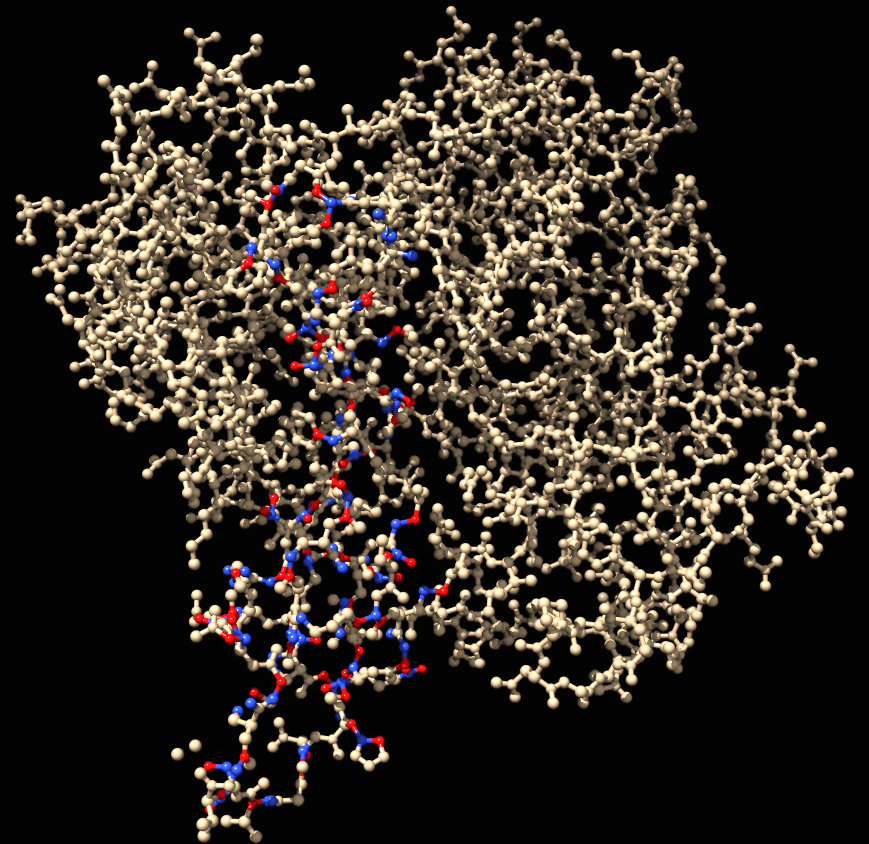}
    \caption{Pattern of excitation of ${\mathcal A}_0$ triggered by excitation of histidine's aromatic ring as shown in Fig.~\ref{aromatic2}. The pattern is recorded at 90th step of evolution.  Resting nodes are light-grey, excited nodes are red, refractory nodes are blue.}
    \label{aromatic3}
\end{figure}

The excited aromatic rings act as generators of excitation in F-actin automata.
Let us consider an example. In Fig.~\ref{aromatic1} we see a histidine's aromatic ring stimulated: one node is assigned excited state and its neighbour refractory state. The wave of excitation travels along the ring clockwise (Fig.~\ref{aromatic2}abc). When excitation reaches a node linked to the rest of the graph the excitation propagates along the `bridge' (Fig.~\ref{aromatic2}d). The excitation then propagates further inside the graph (Fig.~\ref{aromatic2}ef) splitting into two compact excitation patterns at the junction (Fig.~\ref{aromatic2}gh). The overall pattern of excitation in ${\mathcal A}_0$ recorded at 90th step of evolution is shown in Fig.~\ref{aromatic3}.

\section{Discussion}
\label{discussion}

Automaton model of F-actin unit is a fast prototyping tool for studying dynamics of excitation in actin filaments allowing for controlled propagation of localisations at atomic level. Two rules of excitation were analysed. First rule states that a resting node is excited  if it has at least one excited neighbour (${\mathcal A}_0$): this is a classical threshold excitation rule. Second rule states that a resting node is excited if it has exactly one excited neighbour (${\mathcal A}_1$): this may be seen as a rule of non-linear excitation because only narrow band of local activity triggers excitation in the node. We did not consider other ranges of thresholds or excitation intervals, because they always lead to extinction of excitation at the very beginning of  the evolution. Both rules support travelling patterns of excitation. Automata ${\mathcal A}_0$ show longer transient periods, smaller limit cycles and larger average levels of excitation than automata ${\mathcal A}_1$ (Tab.~\ref{tab:statistics}d). When a resting automaton ${\mathcal A}_0$ is stimulated by external excitation of some nodes the excitation patterns spread all over the automaton graph but then activity declines to a global resting state.  
Stimulation of actin automata with a mix of excited and refractory states leads to excitation dynamics with longer transient periods and formation of repeated patterns of excitation, analogous to oscillatory structures.  The limit cycles are stable: an automaton subjected to repeated stimulation always slide back to its pre-stimulation activity level.

\begin{figure}[!tbp]
    \centering
      \includegraphics[width=0.99\linewidth]{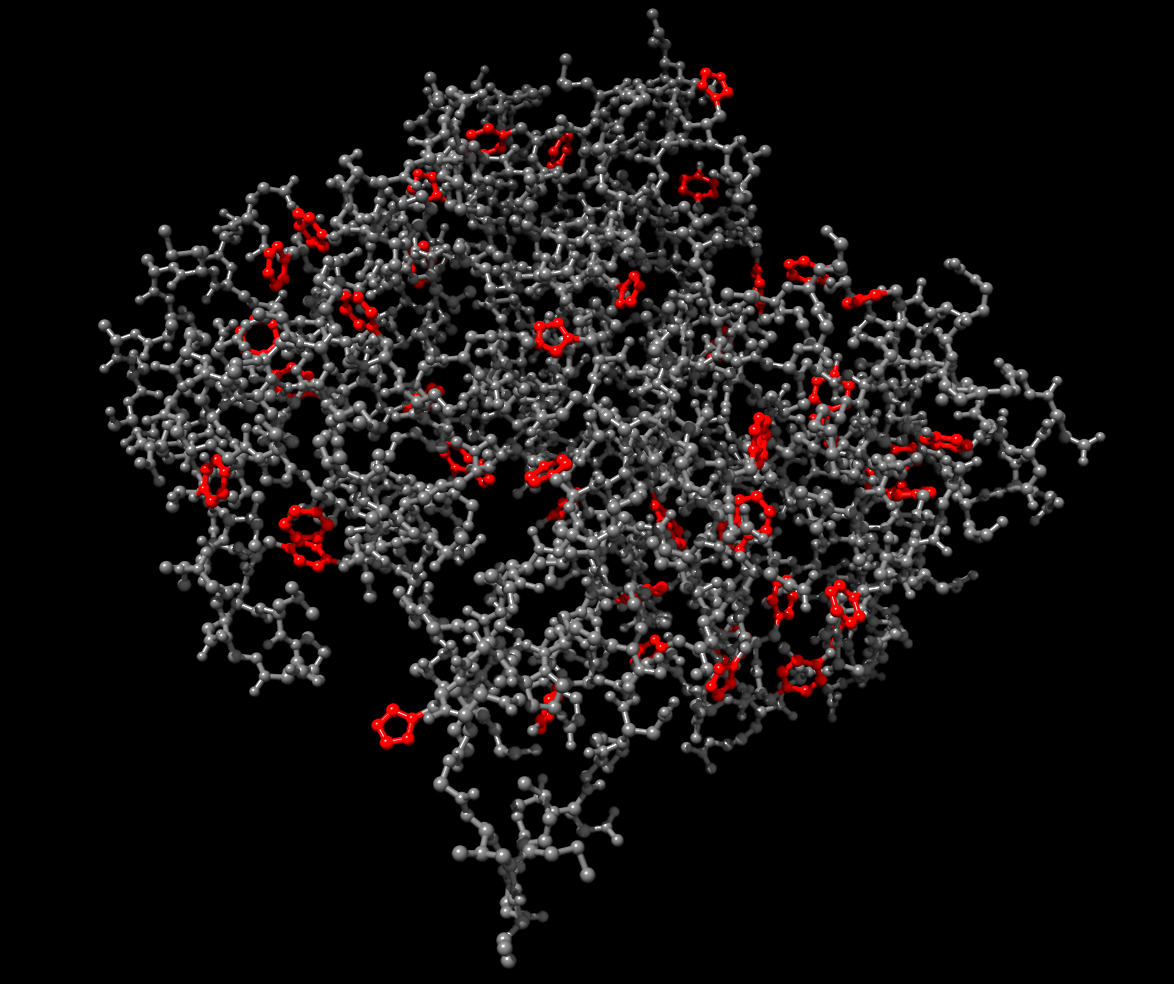}
    \caption{F-actin molecule with aromatic rings highlighted in red.}
    \label{positionsaromatic}
\end{figure}

 Due to substantial noise-tolerance of excitation waves propagating in aromatic rings, the rings could be seen as memory devices in a hypothetical actin computer. Assume excited aromatic ring represents one bit. To write a bit we excite one node and inhibit (force into a refractory state) one of its neighbours. To erase a bit we must excite or inhibit all resting nodes.  An F-actin unit contains 40 rings (8 of histidine, 12 of phenylalanine, 4 of tryptophane, and 16 tyrosine), see  configuration of the aromatic rings in Fig.~\ref{positionsaromatic}. Thus an F-actin unit can store 40 bits. Maximum diameter of an actin filament is 8~nm~\cite{moore1970three,spudich1972regulation}. An actin filament is composed of overlapping units of F-actin (Fig.~\ref{fig:actinballstick}a). Thus, diameter of a single unit is c. 4~nm. Persistent length of F-actin polymer is 17 $\mu$m \cite{gittes1993flexural}, therefore we can assume that it is feasible to write  $32 \cdot 10^4$ bit on a double-strand actin filament. Given appropriate tools to read and write dynamics of excitation in predetermined parts of F-actin molecule we can assume the actin polymer offers us a memory density  64 Petabit per square inch ($6.452\cdot 10^{16}$ per square inch).

\bibliographystyle{plain}
\bibliography{actinbiblio}

\begin{thebibliography}{10}

\bibitem{adamatzky2015actin}
Andrew Adamatzky and Richard Mayne.
\newblock Actin automata: Phenomenology and localizations.
\newblock {\em International Journal of Bifurcation and Chaos}, 25(02):1550030,
  2015.

\bibitem{cingolani2008actin}
Lorenzo~A Cingolani and Yukiko Goda.
\newblock Actin in action: the interplay between the actin cytoskeleton and
  synaptic efficacy.
\newblock {\em Nature Reviews Neuroscience}, 9(5):344--356, 2008.

\bibitem{conrad1996cross}
Michael Conrad.
\newblock Cross-scale information processing in evolution, development and
  intelligence.
\newblock {\em BioSystems}, 38(2):97--109, 1996.

\bibitem{cooper2000cell}
Geoffrey~M Cooper and Robert~E Hausman.
\newblock {\em The cell}.
\newblock Sinauer Associates Sunderland, 2000.

\bibitem{debanne2004information}
Dominique Debanne.
\newblock Information processing in the axon.
\newblock {\em Nature Reviews Neuroscience}, 5(4):304--316, 2004.

\bibitem{dillon2005actin}
Christian Dillon and Yukiko Goda.
\newblock The actin cytoskeleton: integrating form and function at the synapse.
\newblock {\em Annu. Rev. Neurosci.}, 28:25--55, 2005.

\bibitem{fifkova1982cytoplasmic}
Eva Fifkov{\'a} and Rona~J Delay.
\newblock Cytoplasmic actin in neuronal processes as a possible mediator of
  synaptic plasticity.
\newblock {\em The Journal of Cell Biology}, 95(1):345--350, 1982.

\bibitem{gittes1993flexural}
Frederick Gittes, Brian Mickey, Jilda Nettleton, and Jonathon Howard.
\newblock Flexural rigidity of microtubules and actin filaments measured from
  thermal fluctuations in shape.
\newblock {\em Journal of Cell biology}, 120:923--923, 1993.

\bibitem{hameroff1988coherence}
SR~Hameroff.
\newblock Coherence in the cytoskeleton: Implications for biological
  information processing.
\newblock In {\em Biological coherence and response to external stimuli}, pages
  242--265. Springer, 1988.

\bibitem{jaeken2007new}
Laurent Jaeken.
\newblock A new list of functions of the cytoskeleton.
\newblock {\em IUBMB life}, 59(3):127--133, 2007.

\bibitem{kim1999role}
Chong-Hyun Kim and John~E Lisman.
\newblock A role of actin filament in synaptic transmission and long-term
  potentiation.
\newblock {\em The Journal of neuroscience}, 19(11):4314--4324, 1999.

\bibitem{korn1982actin}
Edward~D Korn.
\newblock Actin polymerization and its regulation by proteins from nonmuscle
  cells.
\newblock {\em Physiological Reviews}, 62(2):672--737, 1982.

\bibitem{ludin1993neuronal}
Beat Ludin and Andrew Matus.
\newblock The neuronal cytoskeleton and its role in axonal and dendritic
  plasticity.
\newblock {\em Hippocampus}, 3(S1):61--71, 1993.

\bibitem{moore1970three}
PB~Moore, HE~Huxley, and DJ~DeRosier.
\newblock Three-dimensional reconstruction of f-actin, thin filaments and
  decorated thin filaments.
\newblock {\em Journal of molecular biology}, 50(2):279--292, 1970.

\bibitem{oda2009nature}
Toshiro Oda, Mitsusada Iwasa, Tomoki Aihara, Yuichiro Ma{\'e}da, and Akihiro
  Narita.
\newblock The nature of the globular-to fibrous-actin transition.
\newblock {\em Nature}, 457(7228):441--445, 2009.

\bibitem{priel2006dendritic}
Avner Priel, Jack~A Tuszynski, and Horacion~F Cantiello.
\newblock The dendritic cytoskeleton as a computational device: an hypothesis.
\newblock In {\em The Emerging Physics of Consciousness}, pages 293--325.
  Springer, 2006.

\bibitem{priel2010neural}
Avner Priel, Jack~A Tuszynski, and Nancy~J Woolf.
\newblock Neural cytoskeleton capabilities for learning and memory.
\newblock {\em Journal of biological physics}, 36(1):3--21, 2010.

\bibitem{rasmussen1990computational}
Steen Rasmussen, Hasnain Karampurwala, Rajesh Vaidyanath, Klaus~S Jensen, and
  Stuart Hameroff.
\newblock Computational connectionism within neurons: A model of cytoskeletal
  automata subserving neural networks.
\newblock {\em Physica D: Nonlinear Phenomena}, 42(1):428--449, 1990.

\bibitem{siccardi2015actin}
Stefano Siccardi and Andrew Adamatzky.
\newblock Actin quantum automata: Communication and computation in molecular
  networks.
\newblock {\em Nano Communication Networks}, 6(1):15--27, 2015.

\bibitem{siccardi2016logical}
Stefano Siccardi and Andrew Adamatzky.
\newblock Logical gates implemented by solitons at the junctions between
  one-dimensional lattices.
\newblock {\em International Journal of Bifurcation and Chaos}, 26(06):1650107,
  2016.

\bibitem{siccardi2016quantum}
Stefano Siccardi and Andrew Adamatzky.
\newblock Quantum actin automata and three-valued logics.
\newblock {\em IEEE Journal on Emerging and Selected Topics in Circuits and
  Systems}, 6(1):53--61, 2016.

\bibitem{siccardi2017models}
Stefano Siccardi and Andrew Adamatzky.
\newblock Models of computing on actin filaments.
\newblock In {\em Advances in Unconventional Computing}, pages 309--346.
  Springer, 2017.

\bibitem{siccardi2016boolean}
Stefano Siccardi, Jack~A Tuszynski, and Andrew Adamatzky.
\newblock Boolean gates on actin filaments.
\newblock {\em Physics Letters A}, 380(1):88--97, 2016.

\bibitem{spudich1972regulation}
James~A Spudich, Hugh~E Huxley, and John~T Finch.
\newblock Regulation of skeletal muscle contraction: Ii. structural studies of
  the interaction of the tropomyosin-troponin complex with actin.
\newblock {\em Journal of molecular biology}, 72(3):619--632, 1972.

\bibitem{straub1943actin}
FB~Straub.
\newblock Actin, ii.
\newblock {\em Stud. Inst. Med. Chem. Univ. Szeged}, 3:23--37, 1943.

\bibitem{szent2004early}
Andrew~G Szent-Gy{\"o}rgyi.
\newblock The early history of the biochemistry of muscle contraction.
\newblock {\em The Journal of general physiology}, 123(6):631--641, 2004.

\bibitem{tuszynski1998dielectric}
JA~Tuszynski, JA~Brown, and P~Hawrylak.
\newblock Dielectric polarization, electrical conduction, information
  processing and quantum computation in microtubules. are they plausible?
\newblock {\em Philosophical Transactions -- Royal Soc Series A. Mathematical,
  Physical and Engineering Sciences}, pages 1897--1925, 1998.

\bibitem{tuszynski2004ionic}
JA~Tuszy{\'n}ski, S~Portet, JM~Dixon, C~Luxford, and HF~Cantiello.
\newblock Ionic wave propagation along actin filaments.
\newblock {\em Biophysical journal}, 86(4):1890--1903, 2004.

\end{thebibliography}

\end{document}